\documentclass[%
 reprint,
 amsmath,amssymb,
 aps,
]{revtex4-2}
\usepackage{graphicx}
\usepackage{float}
\usepackage{epstopdf}
\usepackage{color}
\usepackage{amsmath}
\usepackage{amssymb}
\usepackage{wasysym}
\usepackage{etoolbox}
\usepackage{tikz}

\newrobustcmd*{\mycircle}[1]{\tikz{\filldraw[draw=#1,fill=#1] (0,0) circle [radius=0.07cm];}}
\newrobustcmd*{\myholowcircle}[1]{\tikz{\filldraw[draw=#1,fill=white] (0,0) circle [radius=0.07cm];}}
\newrobustcmd*{\myholowsquare}[1]{\tikz{\filldraw[draw=#1,fill=white] (0,0) rectangle ++(4pt,4pt);}}
\newrobustcmd*{\mytriangle}[1]{\tikz{\filldraw[draw=#1,fill=#1] (0,0) --(4pt,0) -- (2pt,4pt);}}
\newrobustcmd*{\myholowtriangle}[1]{\tikz{\filldraw[draw=#1,fill=white] (0,0) --(4pt,0) -- (2pt,4pt);}}
\newrobustcmd*{\mydowntriangle}[1]{\tikz{\filldraw[draw=#1,fill=#1] (-2pt,4pt) --(0pt,0pt) -- (2pt,4pt);}}
\newrobustcmd*{\myholowdowntriangle}[1]{\tikz{\filldraw[draw=#1,fill=white] (-2pt,4pt) --(0pt,0pt) -- (2pt,4pt);}}

\usepackage[colorlinks = true,
            linkcolor = blue,
            urlcolor  = blue,
            citecolor = blue,
            anchorcolor = blue]{hyperref}
\usepackage[pagewise,modulo]{lineno}

\definecolor{lime}{HTML}{A6CE39}
\DeclareRobustCommand{\orcidicon}{%
	\begin{tikzpicture}
	\draw[lime, fill=lime] (0,0) 
	circle [radius=0.16] 
	node[white] {{\fontfamily{qag}\selectfont \tiny ID}};
	\draw[white, fill=white] (-0.0625,0.095) 
	circle [radius=0.007];
	\end{tikzpicture}
	\hspace{-2mm}
}
\foreach \x in {A, ..., Z}{%
	\expandafter\xdef\csname orcid\x\endcsname{\noexpand\href{https://orcid.org/\csname orcidauthor\x\endcsname}{\noexpand\orcidicon}}
}

\begin{document}
	
\title{Elastic and inelastic diffraction of fast neon atoms on a LiF surface}
\author{Maxime Debiossac\orcidC{}}
\author{Peng Pan\orcidA{}}
\author{Philippe Roncin\orcidB{}}
\affiliation{Institut des Sciences Mol\'{e}culaires d'Orsay (ISMO), CNRS, Univ. Paris-Sud, Universit\'{e} Paris-Saclay, F-91405 Orsay, France}
\date{\today}

\pacs{34.35.+a,68.49.Bc,34.50.Cx,79.20.Rf,79.60.Bm,34.20.Cf}

\begin{abstract}
	Grazing incidence fast atom diffraction has mainly been investigated with helium atoms, considered as the best possible choice for surface analysis. This article presents experimental diffraction profiles recorded with neon projectile, between 300 eV and 4 keV kinetic energy with incidence angles $\theta_i$ between 0.3$^\circ$ and 1.5$^\circ$ along three different directions of a LiF(001) crystal surface. These correspond to perpendicular energy ranging from a few meV up to almost 1 eV. A careful analysis of the scattering profile allows us to extract the diffracted intensities even when inelastic effects become so large that most quantum signatures have disappeared. The relevance of this approach is discussed in terms of surface topology. 
		
\end{abstract}

\maketitle

\section{introduction}
Grazing incidence fast atom diffraction (GIFAD) employs the same geometry as reflection high energy electron diffraction (RHEED) and has proven to be a robust technique to track online and in-situ thin films growth on surfaces, in  particular inside a molecular beam epitaxy vessel \cite{Atkinson_2014,Debiossac_PRB_2014,Debiossac_2017}.
Diffraction with neutral atoms originates exclusively from the top most layer~\cite{seifert2010diffraction,Zugarramurdi_SIC}.
It is insensitive to electromagnetic fields, and the absence of the charging effect is well suited for fragile organic layers \cite{seifert_2013alanine,Kalashnyk_2016}.

Elastic diffraction results from the quantum scattering of the projectile on the potential energy landscape (PEL) above the surface. This latter is then accessible by comparison with theory.
For applications, helium is the simplest projectile, behaving as a compact, hardly deformable sphere, so that the projectile is essentially repelled by the surface electronic density $\rho(x,y,z)$ making the interpretation both easier and more valuable in terms of surface engineering (see Ref.~\cite{Winter_PSS_2011,Debiossac_PCCP_2021} for reviews).
With the second-highest binding energy in the periodic table, neon atoms should behave similarly, differing only by a larger number of valence electrons and a larger mass, increasing the momentum transfer to each surface atom and therefore the inelastic effects.
Scattering of neon atoms on a LiF(100) surface has already been published, showing inelastic \cite{Gravielle_2011} and elastic diffraction profiles \cite{Debiossac_JPCL_2020} along the [110] direction together with theoretical analysis.
The present paper reports a full set of diffraction profiles recorded along the [110], [100] and random ([Rnd]) directions with the account of both elastic and inelastic contributions.

Section \ref{ch:setup} presents the experimental arrangement, geometric definitions, and the general strategy for data analysis.
Sections \ref{ch:[110]}, \ref{ch:[Rnd]} and \ref{ch:[100]}  present the diffraction profiles recorded on the Laue circle for neon atoms on the LiF surface oriented along the [110], [Rnd] and [100] directions, respectively. 
Section \ref{ch:discussion} discusses the various strategies used to extract diffracted intensities $I_m$ of each diffraction order $m$ from elastic and inelastic data and their relevance to the PEL.

\begin{figure}
	\includegraphics[width=0.9\linewidth]{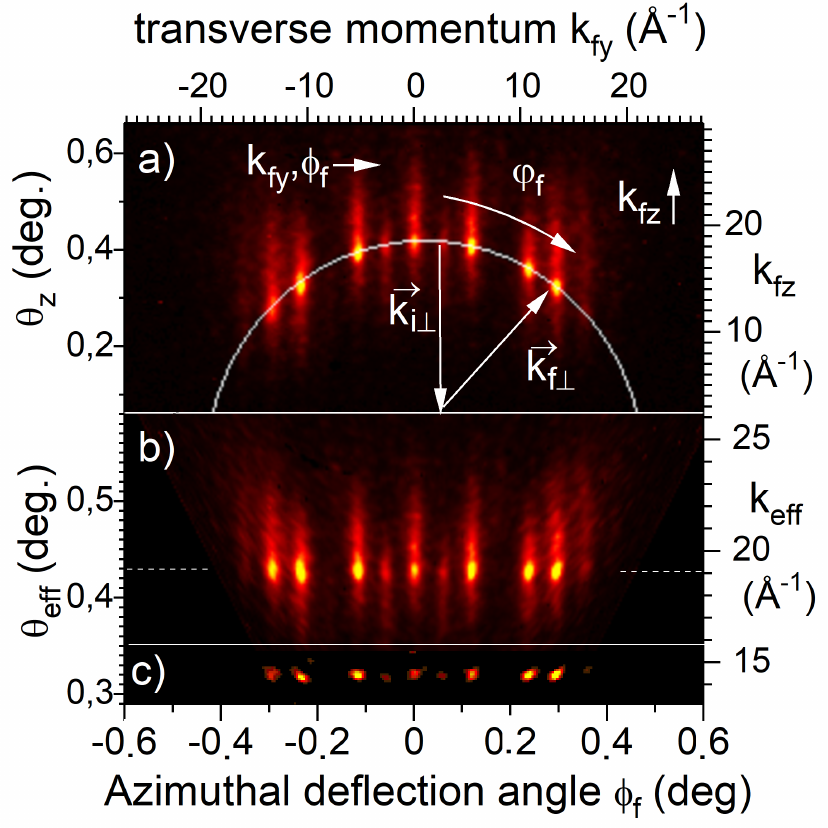}
	\caption{Diffraction pattern recorded for 500 eV neon atoms along LiF[110]. a) Raw scattered intensity $I(k_{fy},k_{fz})$. b) Polar transformed intensity $I(k_{fy},k_\text{eff})$ with the direct beam position as an invariant point and $k_\text{eff}$ the radius of the circle encompassing the direct beam and the diffracted beams (see Ref.~\cite{Lalmi_2012,Debiossac_NIM_2016}). c) Same as b) with a doubly differential filter outlining the elastic component~\cite{Debiossac_NIM_2016}. \label{fig:y13_h221_2D}}
\end{figure}

\begin{figure}
	\includegraphics[width=0.9\linewidth]{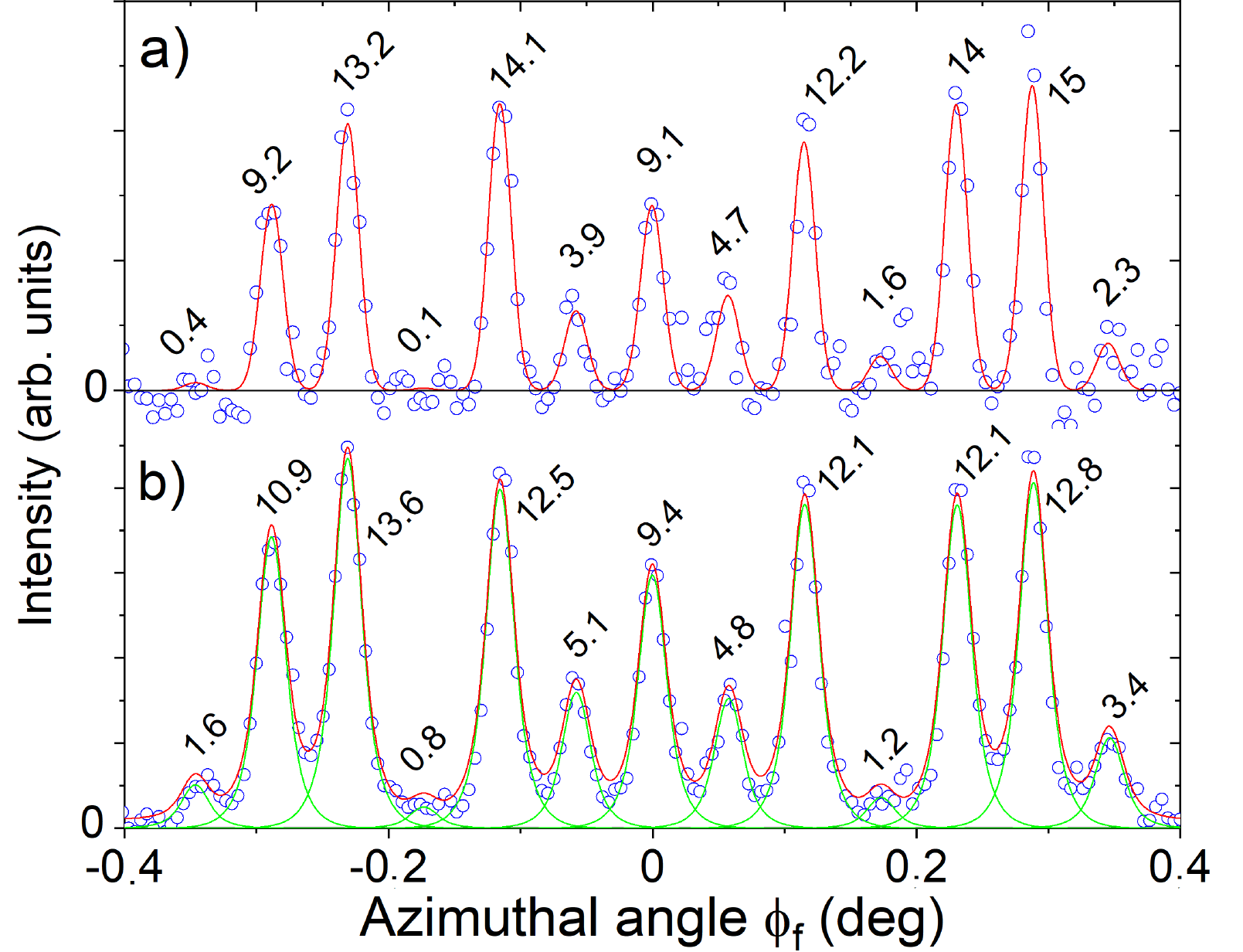}
	\caption{Diffraction profiles from Fig.~\ref{fig:y13_h221_2D}b) and ~\ref{fig:y13_h221_2D}c). Panel a) the data from  Fig.~\ref{fig:y13_h221_2D}c) obtained with a doubly differential filter eliminating inelastic contributions, are fitted by Gaussian functions having a width equal to that of the direct beam $\sigma_b=0.008^\circ$. Panel b) the unfiltered data from Fig.~\ref{fig:y13_h221_2D}b) are fitted by a line profile (green) allowing 60\% of inelastic contribution. Both distributions are taken on the Laue circle, and the intensities $I_m$ (in \%) are reported on each peak. \label{fig:cmp_mex_spe}}
\end{figure}

\section{Setup and data analysis}\label{ch:setup}
The typical arrangement of a GIFAD experiment is detailed in Ref.~\cite{Pan_RSI_2022}, and only a brief description is given here. 
An ion beam in the keV energy range is neutralized by charge exchange and severely collimated before entering a UHV setup where it interacts at grazing incidence $\theta_i \sim 1^\circ$ with a crystalline surface.
The atoms are collected on a position-sensitive detector located $\sim$ 1 m downstream.
\begin{table}
\small
  \caption{Expressions used to describe the diffraction profiles. The * indicates a convolution and $\sigma_b$ is the primary beam resolution. The parameter $a$ is called contrast or visibility of the diffraction peak.
  $A$ is a normalization factor such that $\int_{-\infty}^\infty f(x) dx=1$.}
  \begin{tabular*}{0.48\textwidth}{@{\extracolsep{\fill}}ll}
    \hline
     \hline
Symbol&Formula\\
$L\cdot G_i(\phi) $&$A \cdot e^{-\phi^2/2i^2w^2}/(\phi^2 + w^2/4)$\\
$L\cdot G_i^*(\phi)$&$ [a \delta(\phi) + (1-a) L\cdot G_i(\phi)]*e^{-\phi^2/2\sigma_b^2}$\\
$\phi_{f}$&$\arctan(k_{fy}/k_x)$\\
$\varphi_{f}$&$\arctan(k_{fy}/k_{fz})$\\
$k_\perp$&$(k_y^2 + k_z^2)^{1/2}$\\
$k_\text{eff}$&$\big[\big(\frac{k_{fz}-k_{iz}}{2}\big)^2+k_{fy}^2\big]^{1/2}$\\
$\theta_\text{eff}$&$\arctan(k_\text{eff}/k_x)$\\
    \hline
  \end{tabular*}
\label{Tab:forms}
\end{table}  

In GIFAD, and for well-aligned conditions, the axial surface channeling approximation (ASCA) holds \cite{Zugarramurdi_2012}. 
In this approximation, the motion along the crystal axis, taken as $x$, is decoupled from the one in the perpendicular ($y,z$) plane. 
Labeling the initial and final condition with the subscript $i$ and $f$ respectively, the momentum conservation writes $\vec{k}_f=\vec{k}_i + m.\vec{G}_y + n.\vec{G}_x$ with $G_x=2\pi/a_x$, $G_y=2\pi/a_y$ the reciprocal lattice vectors associated with the surface periodicity $a_x$ and $a_y$ along $x$ and $y$.
Giving an infinite mass to the crystal, the energy conservation writes $|\vec{k}_f|=|\vec{k}_i|$.
In well-aligned GIFAD condition, only the Laue circle is observed, the one corresponding to $n=0$ and corresponding to ASCA, \textit{i.e.} $k_{fy}^2 +k_{fz}^2 = k_{iz}^2$.
It is clearly visible in the raw diffraction pattern shown in Fig.~\ref{fig:y13_h221_2D}a).
After a polar transform~\cite{Debiossac_NIM_2016} bringing the elastic spots on a straight line, the Bragg structure $\phi_f = m \phi_B$ with $\phi_B=\arctan(k_x/G_y)$ is illustrated in Fig.~\ref{fig:y13_h221_2D}b) and Fig.~\ref{fig:cmp_mex_spe}. 
%The surface periodicity $a_x$ and $a_y$ along $x$ and $y$ is associated with reciprocal lattice vectors $G_x=2\pi/a_x$, $G_y=2\pi/a_y$ give rise to momentum conservation $\vec{k}_f=\vec{k}_i + m.\vec{G}_y + n.\vec{G}_x$.
%In well-aligned GIFAD condition, only the Laue circle associated with $n=0$  $k_{fy}^2 +k_{fz}^2 = k_{iz}^2$ is observed, the one corresponding to ASCA and to $k_{fx}=k_{ix}$.
%defining the Laue circle $k_{fy}^2 +k_{fz}^2 = k_{iz}^2$\cite{Laue}.
%This circle is visible in the raw diffraction pattern shown in Fig.~\ref{fig:y13_h221_2D}a) while the location of the elastic spots given by $k_{fy}=m.G_y$ defining the Bragg angle $\phi_B=$ is more clear in the raw diffraction pattern shown in Fig.~\ref{fig:y13_h221_2D}a) and its polar transform \cite{Debiossac_NIM_2016} is reported in Fig.~\ref{fig:y13_h221_2D}b).
Reversely, we assume here that the intensity of the sharp peaks sitting on the Laue circle corresponds to the elastic diffraction. 
This is not fully demonstrated as the final momentum $k_{fx}$ is not measured~\cite{E_loss} but, even under a vacuum of a few 10$^{-10}$ mbar, we observe a drastic reduction of the intensity of these peaks within days, probably due to a progressive reduction of the surface coherence length defined as the mean distance between ad-atoms or defects.

In principle, only the elastic diffraction intensity can be linked directly to the shape of the potential energy landscape.
This later describes the perfectly periodic lattice, which corresponds to surface atoms at their equilibrium position.
Two methods have been proposed to isolate the elastic contribution on the Laue circle.
One assumes that the variation of the inelastic intensity is much slower than the angular resolution of the primary beam $\sigma_b$ so that the application of a doubly differential filter isolates the elastic component.
The filter is a sum of two Gaussians, one positive having a width $\sigma_b$ and the other one negative having a width $2\sigma_b$.
\begin{figure}%[H]
	\includegraphics[width=0.80\linewidth] {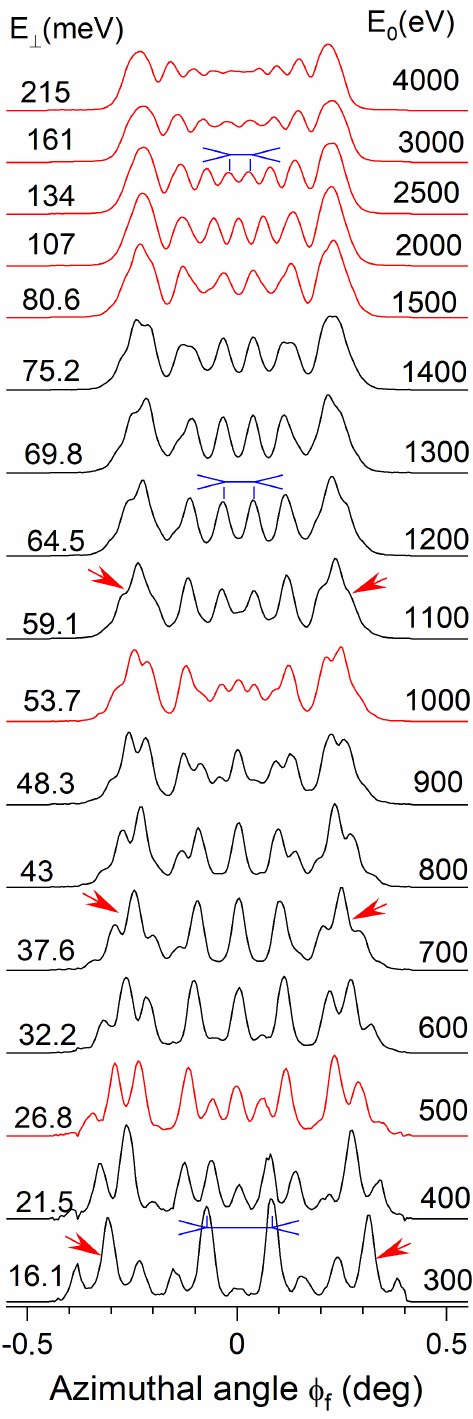}%{Dy_All.eps}
	\caption{Diffraction profiles of neon atoms recorded during an $E$-scan at $\theta_i=0.42^\circ$ on LiF along the [110] direction. The individual peak width appears constant, but their separation $\phi_B$ decreases with $E_0^{-1/2}$. The blue arrows at $E_0=300$, 1200 and 2500 eV point to $m=1$ and $m=-1$ diffraction peaks. The red arrows point to the main rainbow structure made of three to four diffraction peaks progressively merging. \label{fig:Dy_All}}
\end{figure}
An example of a diffraction pattern using such a filter along the z direction is shown in Fig.~\ref{fig:y13_h221_2D}c). 
The 1D profiles with or without application of this filter are reported in Fig.~\ref{fig:cmp_mex_spe}b) and Fig.~\ref{fig:cmp_mex_spe}a) respectively. 

The diffraction pattern in Fig.~\ref{fig:cmp_mex_spe}b) shows quasi-gaussian peaks along the azimuthal direction $k_y$, comforting the initial assumption that elastic peaks shape should be close to that of the primary beam. 
The method is parameter-free, but requires high quality images to limit the noise of the differentiation filter.
%\begin{figure*}
%\centering
%\begin{minipage}{0.4\textwidth}
%  \centering
%  \includegraphics[width=.99\linewidth]{Fit_Energie.pdf}
%  \caption{$E$-scan along [110] at $\theta_i$=0.42$^\circ$, the transverse momentum distribution \myholowcircle{blue} is fitted by $L\cdot G_2$ profile (Table \ref{Tab:forms}). The line-width $\sigma_\phi$ is kept constant at 25 mdeg while the Bragg angle $\phi_B$ scales with $E_0^{-1/2}$.}
%  \label{fig:Fit_energie}
%\end{minipage}%
%\begin{minipage}{.4\textwidth}
%  \centering
%  \includegraphics[width=.99\linewidth]{Fit_Ang.pdf}
%  \caption{$\theta$-scan along [110] at $E_0$=500 eV, Fit of the transverse momentum distribution \myholowcircle{blue} by $L\cdot G_2$ profile (Table \ref{Tab:forms}). The Bragg angle $\phi_B$=0.058$^\circ$ is constant while the line-width $\sigma_\phi$ increases with $\theta_i$.}
%  \label{fig:Fit_Ang}
%\end{minipage}
%\end{figure*}

Without filter, the raw intensity in Fig.~\ref{fig:cmp_mex_spe}a) shows slightly broader peaks with significant wings at their base due to the inelastic component, which may prevent proper measurement of the elastic intensity.
We then rely on an important result established with helium.
It was shown that when elastic diffraction is significant, then the inelastic and elastic relative intensities are identical \cite{Roncin_PRB_2017}.
The challenge is then to find an adequate line-shape describing the combination of elastic and inelastic intensities on the Laue circle.

Our first attempt used an empirical description inspired from numerical simulations \cite{Roncin_PRB_2017} and made of a product of a Gaussian by a Lorentzian having only one width parameter ($L\cdot G_i$ in Table \ref{Tab:forms}).
This strategy was used along the [110] direction and is described in section \ref{ch:[110]}.
Recently, a more general form of the profile \cite{Pan_2023_lateral} was proposed taking into account the resolution and presence of an elastic and inelastic components with relative weight $a$ ($L\cdot G_i^*$ in Table \ref{Tab:forms}).
The parameters $a$ and $w$ of the profiles are measured along the [Rnd] and presented in section \ref{ch:[Rnd]} and used to fit the data recorded along the [100] direction and described in section \ref{ch:[100]}.
All of the diffraction profiles presented here have been recorded on the Laue circle.

\section{[110] direction}\label{ch:[110]}

Two sets of data were recorded with neon atoms along the LiF[110] direction, a $\theta$-scan at $E_0$=500 eV primary energy and an $E$-scan at $\theta_i$=0.42$^\circ$ \cite{Debiossac_JPCL_2020}.

%We focus here on the raw diffraction profiles as displayed in Fig.\ref{fig:Dy_All}.

During an $E$-scan the size of the Laue circle is constant, but the Bragg angle $\phi_B$ indicating the peak separation along $\phi_f$ scales with $E_0^{-1/2}$ so that the number of open channels increases progressively.
%($\phi_B=\arctan k_x/G_y$ where $G_y=2\pi/a_y$ is the reciprocal lattice vector associated with $a_y$, the projected periodicity in the $y$ direction, see e.g.\cite{Debiossac_PCCP_2021}) 
Twenty-eight snapshots were recorded in less than an hour with a resolution $\sigma_b=0.01^\circ$ (0.024$^\circ$ fwhm). Some raw diffraction profiles are reported in Fig.~\ref{fig:Dy_All}.
It shows that line-shape does not increase significantly, but the narrowing of the peak separation leads to a progressive merging of the peaks to a smooth quasi-continuous profile where the distinction between diffraction peaks and supernumerary rainbows~\cite{Winter_2005rainbow,Debiossac_PCCP_2021} is unclear.
%The outer or principal rainbows correspond the existence of an impact parameter $y$ such that the deflection angle $\varphi$ is maximum. As the energy increases, the rainbow structure is made of an increasing number of diffraction peaks corresponding to trajectories with only a moderate phase difference.Starting from the outer rainbow, the $n^{th}$ supernumerary rainbows are the locations where the phase difference are $n\pi}$. Since the corresponding trajectories are more separated, the phase difference evolves faster and faster making less diffraction peaks to contribute
\begin{figure}\centering
\includegraphics[width=0.8\linewidth]{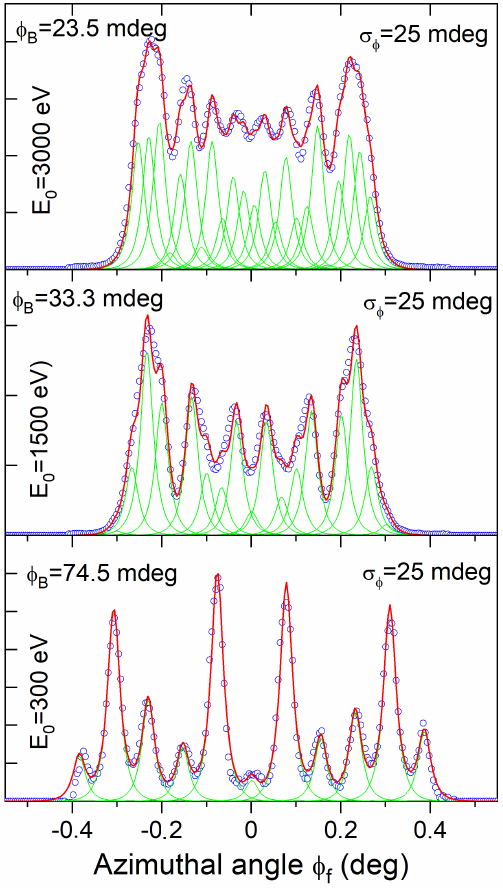}
\caption{$E$-scan for neon atoms along the [110] of LiF at $\theta_i=0.42^\circ$. The transverse momentum distribution \myholowcircle{blue} is fitted by $L\cdot G_2$ profile (green, Table \ref{Tab:forms}). A reasonable fit is obtained with a constant line-width $\sigma_\phi = 25$ mdeg while the Bragg angle $\phi_B$ scales with $E_0^{-1/2}$. \label{fig:Fit_energie}}
\end{figure}

This continuous profile gives the impression that the semi-classical link between topology and diffraction profile is preserved, but the progressive weakening of the contrast in the center indicates clearly that the information on the corrugation amplitude is degraded.
This is in part balanced by the fact that the natural measurement unit is the wavelength $\lambda_\perp$, and this latter reduces at larger values of $E_\perp$.
It should be noted that this presentation of the scattering profiles is more quantitative than the color plots of the diffraction chart~\cite{Winter_PSS_2011,Debiossac_JPCL_2020}, which mainly gives an overall impression of where the maxima are located, leaving weak contributions and line-shapes hardly visible.
The diffracted intensities $I_m$ are derived by a fit were all diffraction orders have identical $L\cdot G_2$ line-shapes as shown in Fig. ~\ref{fig:Fit_energie}.
The diffraction profiles are reasonably well fitted using the same $L\cdot G_2$ line-shape with a width parameter $w$ such that the standard deviation is $\sigma_\phi=25$ mdeg, hardly more than the primary beam $\sigma_b=9$ mdeg.
The topmost image where $\phi_B$ has become half of the linewidth (\textit{i.e.} less than the fwhm) suggests that unless the exact target alignment is known with an accuracy better that a fraction of the Bragg angle $\phi_B$, trying to recover the exact intensity $I_m$ of contributing peaks is a daunting task.

\begin{figure} \centering
\includegraphics[width=0.8\linewidth]{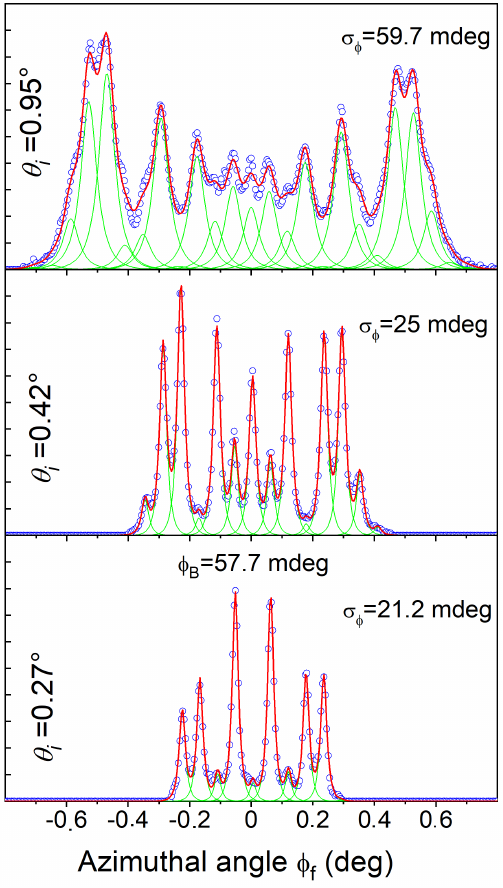}
\caption{$\theta$-scan for neon atoms along LiF[110] at $E_0=500$ eV. Fit of the transverse momentum distribution \myholowcircle{blue} by $L\cdot G_2$ profile (green, Table \ref{Tab:forms}). The Bragg angle $\phi_B=0.058^\circ$ is constant while the line-width $\sigma_\phi$ increases with $\theta_i$. \label{fig:Fit_Ang}}
\end{figure}

During a $\theta$-scan the primary beam energy is constant so that the Bragg angle $\phi_B$ is essentially constant ($\cos\theta_i\sim 1$ within a few $10^{-4}$) while the radius $\theta_i$ of the Laue circle increases allowing also more and more diffraction orders to contribute to the diffraction pattern.
The $\theta$-scan consists in 45 diffraction images recorded between 0.27$^\circ$ and 0.94$^\circ$ with a resolution of $\sigma_b=8$ mdeg. Three diffraction profiles recorded on the Laue circle are displayed in Fig.~\ref{fig:Fit_Ang}.
Both the number of contributing diffraction orders and the line-width progressively increase with $\theta_i$, reducing the visibility of the peaks and the contrast in the center. 
Both the $E$-scan and the $\theta$-scan yield very similar intensities $I_m$ when plotted as a function of the energy $E_\perp$ \cite{Debiossac_JPCL_2020} confirming, once more~\cite{momeni2010grazing,Winter_PSS_2011}, the validity of ASCA.

%Fig.\ref{fig:DWF_Ep_Thetascan} reports the evolution of the Debye-Waller factor extracted from the polar scattering profiles (see inset of Fig.\ref{fig:DWF_Ep_Thetascan} and Ref.\cite{Debiossac_JPCL_2020} and \cite{Pan_2021_polar}).
\begin{figure} \includegraphics[width=0.9\linewidth]{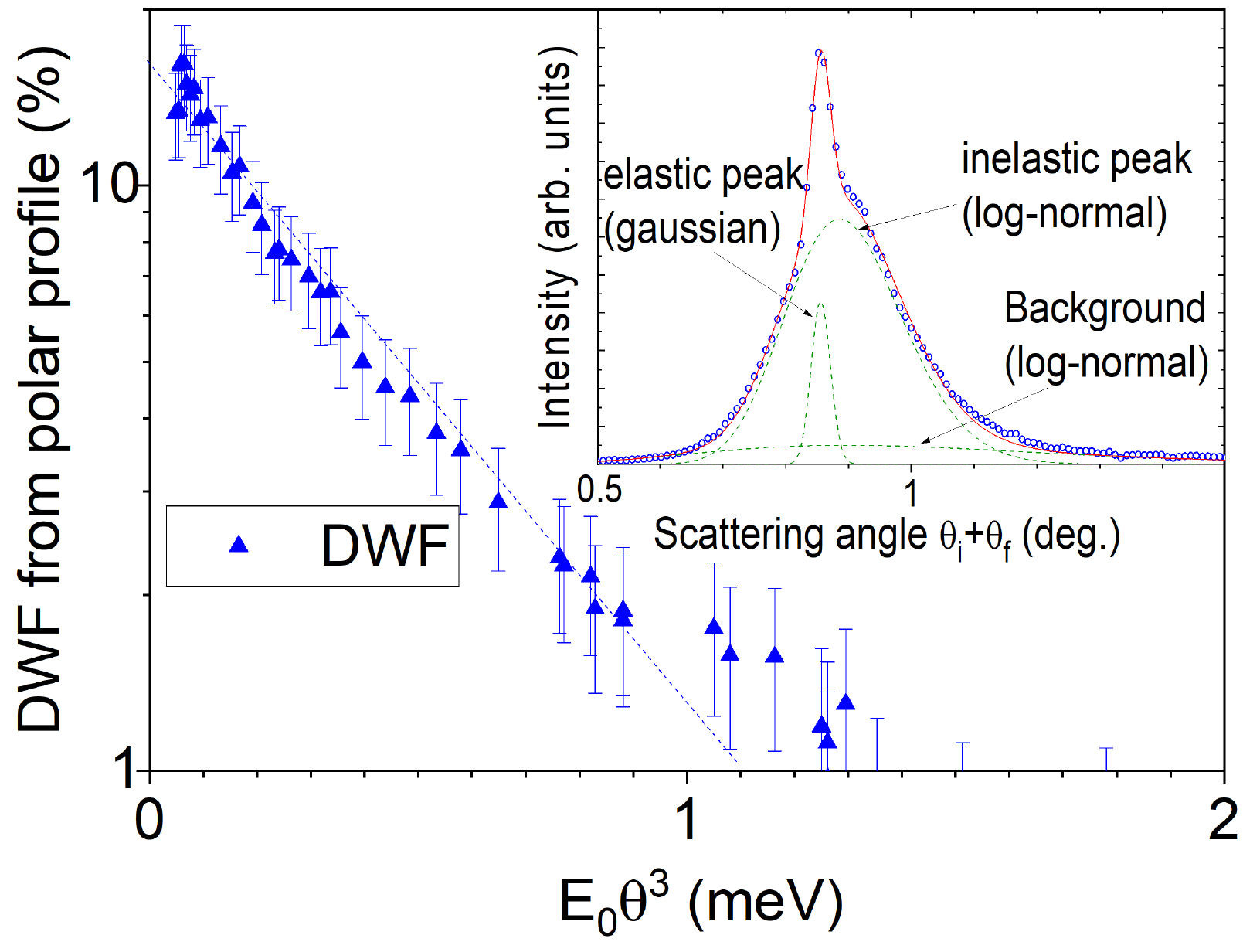}
\caption{Debye-Waller factor for 500 eV neon atoms along LiF[110] versus $E_0\theta^3$. The elastic ratio extracted from the polar profile is displayed with blue triangles (\mytriangle{blue}). The surface is at room temperature. 
The inset corresponds to $\theta_i$=0.42$^\circ$. The dashed blue line is a guide for the eyes.\label{fig:DWF_Ep_Thetascan}}
\end{figure}

%The leveling of $\sigma_\phi$ at low values of $E_\perp$ is robust, however its exact value is difficult to interpret because a unique $L\cdot G_2$ line profile was used to described both elastic and inelastic contribution and also because the contribution of the primary beam is not subtracted (quadratically).
%However, the use of a single $L\cdot G_2$ form for a variable weight of elastic and inelastic contribution does not allows a proper measurement of the the value of the width is difficult to interpret.
%We did not plot the evolution of the linewidth because, when using a single $L\cdot G_2$ form for a variable weigth of elastic and inelastic contribution the value of the width is difficult to interpret.
%This motivate in part the development of more adequate line profiles $L\cdot G_1^*$ profiles (Table \ref{Tab:forms} and Ref.\cite{Pan_2023_lateral}).
In Fig.\ref{fig:DWF_Ep_Thetascan}, we report the Debye-Waller factor (DWF) for neon atoms at 500 eV. The DWF is estimated here as the fraction of the elastic peak in the observed polar scattering profile, as detailed in Ref.~\cite{Pan_2021_polar} and shown in the inset of Fig.~\ref{fig:DWF_Ep_Thetascan}.
This technique does not allow for a relevant estimate of the DWF below 1\%, in part because the log-normal profile is not a perfect description \cite{Roncin_PRB_2017}, it is only the best available so far. %The error bar is difficult to estimate because it is highly sensitive to possible broadening at the base of the elastic peak (see \textit{e.g.}\cite{Pan_2021_polar}).
The DWF is plotted as a function of $E_0\theta^3$, as suggested in Refs.~\cite{Manson_PRB_2008,Rousseau_2008,Roncin_PRB_2017} because this value corresponds to the classical energy loss in the multiple successive grazing collision regime of GIFAD.
The quantity $E_0\theta^3$ is expected to replace the binary recoil energy appearing in the standard DWF used in single scattering conditions such as X-ray, neutrons or thermal energy helium diffraction.
In GIFAD, for helium projectile impinging on a LiF surface at room temperature, this 1\% value of the DWF is reached with $E_0\theta^3 \sim 8-10$ meV~\cite{Pan_temp_2022} while for the case of neon atoms, it is reached here between 1.1 and 1.3 meV. 
This is consistent with the fact that for comparable trajectory, all the binary classical energy transfers to the surface atoms should scale with the projectile mass.

More important for the analysis in the next sections, the fraction of elastic scattering on the Laue circle, and called contrast or visibility, is given by the relative height of the elastic Gaussian peak at its maximum. 
It is typically $\sim\sigma_\theta/\sigma_b$ larger than the DWF.
In the example in the inset of Fig.~\ref{fig:DWF_Ep_Thetascan}, the DWF is only 10\% while the contrast $a$ (see $L\cdot G_i^*$ in Table \ref{Tab:forms}) at the specular position is close to 41\%.

%\begin{figure} \includegraphics[width=0.95\linewidth]{W_DWF_Eperp.pdf}
%\caption{$\theta$-scan at 500 eV, the elastic scattering ratio \mytriangle{blue} measured from the polar profile is plotted in log scale together with the azimuthal line-width \myholowcircle{blue} of the $LG_2$ profile ignoring the balance of elastic and inelastic contributions. Lines are to guide the eyes.\label{fig:DWF_Ep_Thetascan}}
%\end{figure}

\section{[Rnd] direction}\label{ch:[Rnd]}
The [Rnd] and [100] directions were recorded more recently with 1 keV neon atoms and a new detector \cite{Lupone_2018}.
The strategy to analyze the diffraction pattern has also evolved.
Following Ref.~\cite{Pan_2023_lateral}, we use a $L\cdot G_1^*$ line-shape (Table \ref{Tab:forms}) which takes explicitly into account the primary beam resolution $\sigma_b$ (here also of 9 mdeg) and the elastic ratio $a$ on the Laue circle so that the measured width parameter $w$ corresponds to that of the inelastic component only.

\begin{figure}
\includegraphics[width=0.9\linewidth]{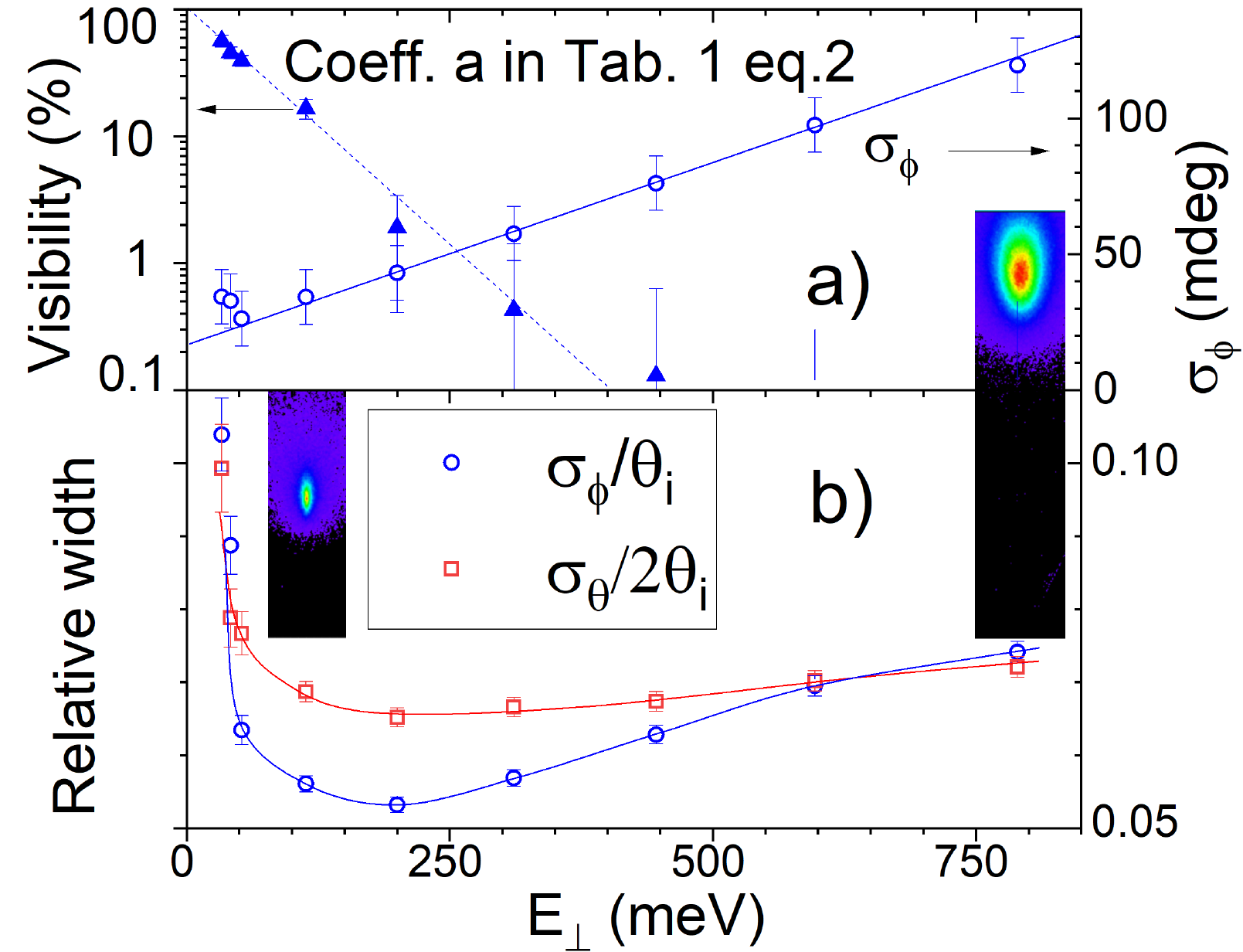}
\caption{Scattering of 1 keV Ne along a [Rnd] direction of LiF. a) Azimuthal width (\myholowcircle{blue}) and visibility (\mytriangle{blue}) both measured on the Laue circle. 
 b) Relative azimutal $\sigma_\phi /\theta_i$ (\myholowcircle{blue}) and polar $\sigma_\theta /2 \theta_i$ (\myholowsquare{red}) widths. 
 The two insets are the 2D scattering patterns at $\theta_i$ = 0.6$^\circ$ and 1.6$^\circ$ respectively. At the highest investigated energy of 4 keV and $\theta_i=4^\circ$, the perpendicular energy is $E_\perp$=19.5 eV and the standard deviation is $\sigma_\phi \sim $
  0.58$^\circ$, ten times larger than $\phi_B$ and $\sigma_\theta \sim$ 1.2 $\sigma_\phi$. In a) and b) lines are guides to the eyes.
\label{fig:Ne_1keV_Rnd}}
\end{figure}

In the quantum regime, a random direction can be defined as a direction where only the specular peak ($m=0$) is present, indicating a surface seen as perfectly flat so that no structural information can be extracted (see \textit{e.g.} Ref.~\cite{Zugarramurdi_2012}). In the present case, the exact direction was 10$^\circ$ away from the [100] direction. 2D images of the scattering profiles are displayed as insets in Fig.~\ref{fig:Ne_1keV_Rnd}. 

The dynamic information, which is the ratio of elastic to the total scattering intensity, identified here to the DWF and the inelastic polar ($\sigma_\theta$) and azimuthal ($\sigma_\phi$) profiles are comparatively easy to measure and are reported in Fig.~\ref{fig:Ne_1keV_Rnd}a).
The lines drawn to guide the eyes indicate that the visibility on the Laue circle drops by four orders of magnitude between an (extrapolated) incident perpendicular energy of 0 meV and $\sim$ 400 meV perpendicular energy.
The inelastic lateral width increases linearly above 100 meV following approximately the equation $\sigma_\phi\sim$ 16 + 135 mdeg/eV.
With helium and under restricted conditions, these polar and azimuthal profiles were identical, irrespective of the surface orientation \cite{Pan_2021_polar,Seifert_2015,Pan_2023_lateral}.

\section{[100] direction}\label{ch:[100]}
\begin{figure}
	\includegraphics[width=0.8\linewidth]{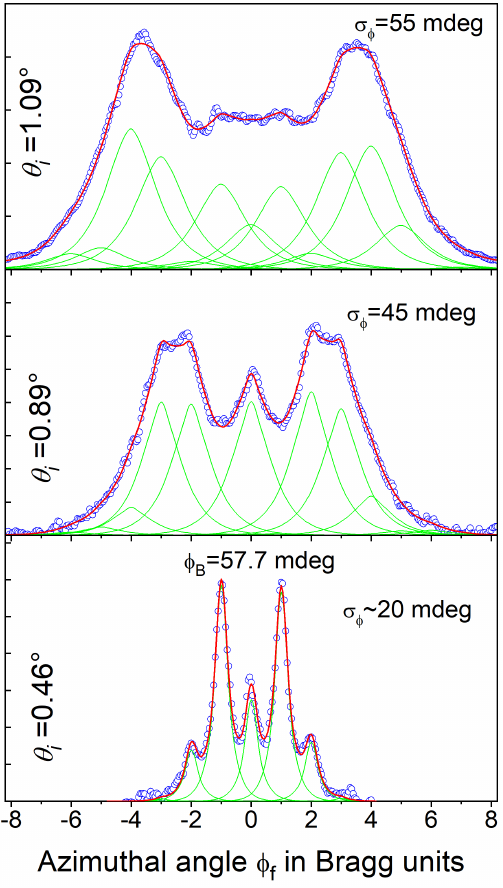}
	\caption{Scattering profiles of 1 keV neon along the LiF[100] direction at $\theta_i=0.46^\circ$, 0.89$^\circ$ and 1.09$^\circ$ ($E_\perp=65$, 240 and 363 meV respectively) fitted with a $L\cdot G_1^*$ profiles using the width and visibility parameters from the [Rnd] direction in Fig.~\ref{fig:Ne_1keV_Rnd}.  \label{fig:Ne[100]_fit_example}}
\end{figure}
\begin{figure}
\includegraphics[width=0.9\linewidth]{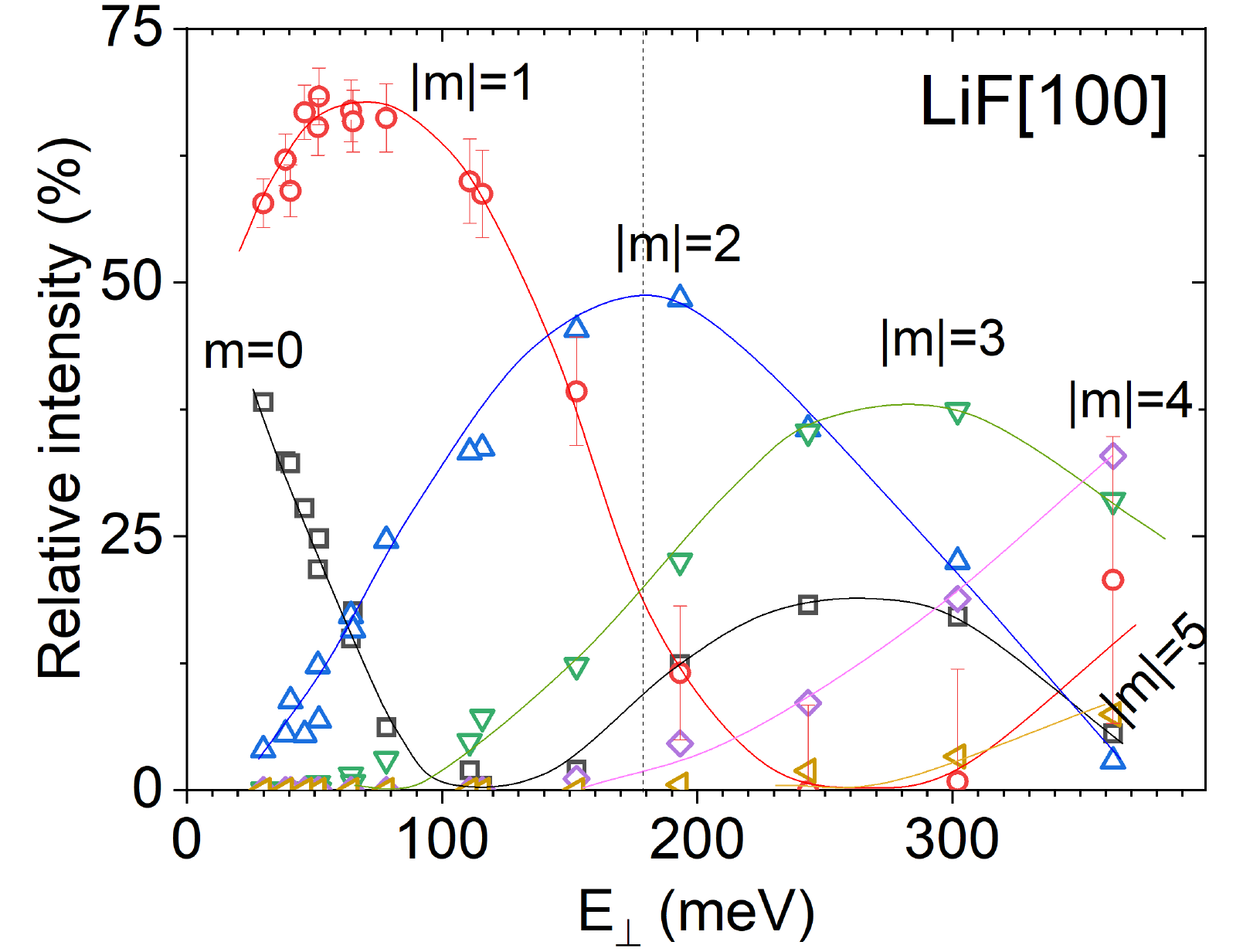}
\caption{Intensities $I_m$ of the diffraction orders $+m$ and $-m$ for 1 keV neon along the LiF[100] direction. The vertical dotted line indicates the region beyond which elastic contribution is less than 1\% and inelastic line-width (fwhm) larger than $\phi_B$, making the fit unstable and almost impossible without a proper line-shape.  Lines are only a guide to the eyes and typical error bars are reported for the m=$\pm$1 order. \label{fig:Ne_1k_100_Im_Ep}} 
\end{figure}

The diffraction profiles recorded in a $\theta$-scan at 1 keV were fitted by $L\cdot G_1^*$ lineshapes with 
$a$ and $w$ parameter taken from Fig.~\ref{fig:Ne_1keV_Rnd}a) measured along the [Rnd] direction at same energy $E_0$.
There is obviously not enough data points in Fig.~\ref{fig:Ne_1keV_Rnd}a) at low values of $E_\perp$ so that comparatively large values were selected in Fig.~\ref{fig:Ne[100]_fit_example}.
The only parameter of the fits are the set of relative intensities $I_m$ but the quality appears reasonable, and the evolution of $I_m$ with $E_\perp$ reported in Fig.~\ref{fig:Ne_1k_100_Im_Ep} resembles a typical evolution of Bessel function as expected from simple cosine corrugation function. Note that in Fig.~\ref{fig:Ne_1k_100_Im_Ep} the intensity $I_m$ of the diffraction orders $m=\pm 1$ recorded at $\theta_i=0.89^\circ$ are so low that they are not visible whereas the associated energy $E_\perp$ of 240 meV is such that the elastic contribution on the Laue circle is only a few $\permil$. It does not prove that the inelastic signal is fully coherent, but the coherent description with a lineshape broadened by inelastic effects seems to be a fair description of the progressive weakening of visible quantum features.

The vertical dotted line indicates the region beyond which the elastic contribution is less than 1\%. Below this line, a self-consistent evaluation of $I_m$ with internally optimized diffraction profile would probably be possible. However,  above this limit, a direct evaluation of the intensities $I_m$ without \textit{a priori} information on the linewidth would have been questionable. It should also be outlined that at comparatively large incidence angle, the polar scattering profile becomes so large that selecting a narrow slice having a width close to that of the primary beam $\sigma_b$ corresponds to a weak fraction of the intensity. In Fig.~\ref{fig:Ne[100]_fit_example} the profile recorded at $\theta_i$=1.09$^\circ$ corresponds to only $\sim$~5\% of the scattered intensity because the fully inelastic polar profile is $\sim$~0.4$^\circ$ wide (fwhm).
The size of the polar slice selected for the profile could be enlarged beyond $\sigma_b$ because no elastic diffraction is visible, but not too much otherwise the contribution of different effective angles $\theta_\text{eff}$ would smear the rapid evolution of $I_m$. 
In particular, the ability to measure a weak value of a given diffraction order would be compromised by a simple averaging effect.
We should also stress that the independence of the lateral profile to the surface orientation cannot be demonstrated beyond this line where individual peaks are not resolved.
One can, at most, confirm that the profile measured along a [Rnd] direction offers a decent fit along [100] or [110] as illustrated below in Fig.\ref{fig:Ne[100]_fit_example}.
However, this is not fully convincing evidence that the lineshapes are indeed identical since many overlapping lines contribute to a diffraction profile which becomes increasingly smooth as visible in top profiles in Fig.~\ref{fig:Dy_All}.

\section{Discussion}	\label{ch:discussion}	
In this section, we assume that the inelastic component originates only from the interaction with surface phonons~\cite{Taleb2017multiphonon,tamtogl_2013} along the projectile trajectory above the surface. Compared with helium, the large mass of neon strongly enhances the inelastic effect, so that its contribution cannot be neglected.
Assuming that inelastic and elastic intensity ratios are identical on the Laue circle and that the lineshapes are identical for all surface orientations, we have produced diffracted intensities $I_m$ in conditions where no elastic diffraction is visible.
We have no guarantee that these intensities are correct because the two starting assumptions were derived only in conditions where elastic diffraction is significant.
The question can be turned differently, are the intensities $I_m$ derived from elastic and inelastic intensities always comparable or is this valid in a limited range?

This naive question has been present since the beginning of GIFAD.
The elastic component corresponds to a surface with atoms fixed at equilibrium positions because this is the only configuration that offers perfectly periodic conditions.
How is this idealized description connected with an actual surface where the atoms have a mass and undergo thermal motion?
The answer is that the surface is not classical but quantum. 
The vibrations at surfaces are specific phonon modes that can be further simplified as independent harmonic oscillators using the Debye Model.
The probability $p_e$ to leave a harmonic oscillator in its ground state, with wavefunction $|\psi\rangle$ in response to a sudden momentum transfer $\delta k$ is $p_e=|\langle \psi|e^{i\delta k}|\psi\rangle|^2$, known as the Lamb-Dicke probability or recoilless emission (or absorption) probability in atomic spectroscopy or in the cold atom community~\cite{wineland1992sisyphus,jurczak_1996atomic}.
If no excitation occurs, the scattering takes place from the center of the vibration wave-function $z_e=\langle \psi|z|\psi\rangle$ which is indeed sitting at the equilibrium position in spite of a Gaussian thermal position distribution of variance $\sigma_z^2(T)=\langle \psi|z^2|\psi\rangle$ that never reaches 0.

The question is now what do we see in inelastic diffraction?
Starting with the observation that, when elastic diffraction is present and within a limited range around the Laue circle (typically $|\theta_f-\theta_i| \> \sigma_\theta$), the structural information $I_m$ derived from the inelastic profiles compares with the elastic one.
Of course, an effective wavelength has to be defined but it was shown~\cite{Debiossac_NIM_2016,Roncin_2018} that using $k_\text{eff}$ (Table \ref{Tab:forms}) provides intensities $I_m$ comparable with the one derived from elastic diffraction at $\theta_i=\theta_\text{eff}$.
A model was developed where inelastic diffraction is seen as a perturbation of the elastic trajectory. Among the N lattice sites encountered, each binary collision can turn inelastic (with a probability $p_e$), allowing a derivation of the DWF adapted to GIFAD~\cite{Rousseau_2008,Manson_PRB_2008} and essentially confirmed by experiment at different temperatures~\cite{Pan_temp_2022}. 
On the other hand, the idea that the scattering profile could be decomposed in terms of the number of inelastic binary collisions, each contributing to a finite angular broadening $\delta\theta$ was invalidated. 
%$\delta\theta$ is calculated in a classical binary collision model with a simplified binary interaction potential is used together with the atomic mass of the surface atom and its thermal amplitude $\sigma_z^2(T)$.
Even at the lowest investigated angles, the agreement with the experimental inelastic polar profiles was reached~\cite{Pan_2021_polar} only with the classical limit where $\sigma_\theta^2=N\delta_\theta^2$ as if a single inelastic collision would turn all $N$ other collisions inelastic, as considered in case B of Ref.~\cite{Manson_PRB_2008}.

%There have been attempts to expend the polar inelastic profile in terms of the number $n$ of individual inelastic events were each one would contribute to a broadening of the final polar profile easily calculated in a classical binary collision model where a simplified binary interaction potential is used together with the atomic mass of the surface atom and its thermal amplitude $\sigma_z^2(T)=\langle \psi|z^2|\psi\rangle$.
%This was not successful and even in conditions were the probability of an inelastic event is small, the inelastic profile appeared to be better modelled by a classical limit where all binary collisions along the trajectories had turned inelastic \cite{Pan_2021_polar}.
For temperatures below the Debye temperature, this should correspond to scattering from a surface where the surface atoms adopt a classical position distribution. In other words, the inelastic polar scattering distribution appears close to that arising from classical mechanics.
In this case the intensities $I_m$ derived from inelastic conditions should be related to the thermally averaged surface, not the one at the equilibrium position. 
These are different in GIFAD because the diffraction takes place along the well-aligned rows on the effective 2D potential $V_{2D}(y,z)=(1/a_x)\int_0^{a_x}V_{3D}(x,y,z) dx$ with $a_x$ any multiple of the lattice parameter along $x$. 
Whether explicit~\cite{Rousseau_2007} in a 2D calculation or implicit using a 3D trajectory~\cite{Diaz_2016b}, the integral or average in the effective 2D potential is not the same with surface atoms thermally displaced or at equilibrium position because of the exponential character of the potential along $z$. 
One is temperature dependent while the ideal one measured in elastic diffraction is not. This is visible in Fig. 6 of Ref.~\cite{Pan_temp_2022} where $I_m$ recorded on the Laue circle in quasi elastic conditions at different temperatures fall on top of each other.

On the theoretical side, the rumpling of the Li$^+$ ions was estimated by comparing with experimental values of $I_m$. The value derived from an ideal surface model is $\simeq$ 20 \% less than that estimated from the thermally averaged surface~\cite{Schuller_rumpling}.
Also in Ref.~\cite{gravielle2020phonon}, depending on the surface representation used, the intensity calculated at the rainbow angle can vary by a factor of three to four, and some intensities $I_m$ can switch from intense to negligible \cite{frisco_2023_decoherent}.
These calculations are presented as elastic, but no sharp peak~\cite{E_loss} is present and the scattering profiles compare with the one called here inelastic, possibly because the Lamb-Dicke effect is not properly taken into account.
In addition, the inelastic effects are probably underestimated because the author introduces a restriction in the thermal amplitude~\cite{frisco_2019_phonon,frisco_2020_thermal,Diaz_2022} that is supposed to guarantee an elastic scattering.

%The problem is also on the experimental side, some authors~\cite{Bocan_2020} claim, without restriction nor any reference, that there is no difference between the intensities $I_m$ elastic and inelastic diffraction profiles. 
%In addition only 2D color plot highly saturated in color so that no detailed comparison of diffracted intensities is possible.
%The experimental data do not show any elastic component, and the theory is presented as elastic. Both theory and experiment are shown in 2D color plots highly saturated in color so that no detailed comparison of diffracted intensities or even location of maxima is possible.Hence the quote in Ref.~\cite{Bocan_2021} of an “excellent agreement with the experiment” is probably excessive. In addition a corrugation amplitude is extracted from the data and called anomalous because it increases at low value of the impact energy $E_\perp$. 
%This situation is known since fifty years and is most simply fixed by using the Beeby correction replacing the impact energy $E_\perp$ by the effective impact energy $E_\perp+D$, where $D$ is the well-depth due to the attractive forces.
%When used in the hard wall formula, it describes the refraction effects that increases the rainbow angle (see e.g.  Eq.2.8 of Ref.~\cite{Farias1998} or Ref.~\cite{Debiossac_PRB_2016} in GIFAD). 
%Maybe some anomalous behavior is present, but the analysis presented has significant weaknesses and some are discussed in a recent approach using second-order perturbation theory~\cite{allison2022perturbation}.

%Returning to the surface representation, 
It seems clear that, from the theoretical side, there are significant differences between the intensities $I_m$ derived from the ideal and the thermally averaged surfaces.
Can the second one be associated with the experimental inelastic diffraction?
The answer is probably yes when elastic diffraction has disappeared, but the answer could be progressive.
The equivalence between $I_m$ from elastic and inelastic diffraction observed when both are present~\cite{Roncin_PRB_2017} appears in contradiction with the strict association inelastic-thermally averaged.
These facts could be re-investigated in more detail, but a strong difference between elastic and inelastic values of $I_m$ would probably have been detected.
The gradual evolution is also supported by the recent finding~\cite{Pan_2023_lateral} that at low values of $E\theta^3$, where elastic diffraction is important, the polar angle dependence of the azimuthal inelastic width $\sigma_\phi$ is linear with a minimum at the specular angle: $\sigma_\phi (\theta_f)=\sigma_{\phi_s} +\alpha |\theta_f-\theta_i|$ where $\sigma_{\phi_s}$ is the width at the specular angle $\theta_f=\theta_i$.
This could support a perturbative approach explaining that the elastic and inelastic intensities $I_m$ are identical on the Laue circle.
However, this dependence $\sigma_\phi (\theta_f)$ rapidly becomes more complex progressively loosing memory of the specular position~\cite{Pan_2023_lateral}.
This could be a sign of inelastic diffraction starting to probe the thermally averaged surface, \textit{i.e.} the one with classically distributed atoms.
How fast or how progressive is the transition remains to be investigated both theoretically and experimentally, but no convincing model is available yet.
It could be that inelastic effects start with long wavelength phonons, as suggested in a calculation trying to model a quantum surface \cite{Schram_2018}, but no scattering distribution was presented. 

It should also be mentioned that the inelastic component may have various origins, we have focused here on the phonon contribution but, most likely, a limited surface coherence is responsible for the absence of elastic diffraction in many experiments, including our first publications \cite{Rousseau_2007,Rousseau_2008,momeni2010grazing} and may also contribute partly as suggested by the fact that the DWF seems to saturate at the lowest values of $\theta_i$~\cite{Pan_temp_2022}.
The question of the nature of the defects limiting this surface coherence is difficult and probably important, ad-atoms may have different consequences on the inelastic profiles than missing atoms.

Note also that when important electronic excitations are present, such as LiF excitons~\cite{Roncin_1999} with an energy above 10 eV are excited, diffraction disappears \cite{lienemann2011coherence}.
For weaker electronic excitation such as the one at the Fermi level of a metal, the momentum exchanged could be less than a reciprocal lattice vector, allowing a contribution from electron system \cite{bundaleski_2011} in inelastic diffraction.

\section{Summary and Conclusion}	
We have explored the line profile and DWF with neon projectile for which the inelastic effect are significantly larger than for helium. 
We have extracted diffracted intensities $I_m$ over a broad range of impact energy $E_\perp$ using the azimuthal profile recorded along a [Rnd] direction as a reference.
From the point of view of data analysis, this seems to be an efficient technique however, when the profile becomes too broad, the uniqueness of the deconvolution is far from guaranteed, in particular in the quasi-specular region ($m\sim0$) where the oscillation of intensities $I_m$ with $E_\perp$ tend to be $\pi$ shifted~\cite{Debiossac_PRB_2014,Debiossac_PCCP_2021}.
In this case, it would be more reasonable to provide an analytic form of the line shape so that comparison with independent theoretical predictions is possible after convolution by this form.
The experiments should then provide both the diffraction profiles along a crystal axis and a reference profile recorded along the [Rnd] direction.
This approach could also be used when the primary beam is slightly misaligned with the crystal axis.
It is comparatively easy for the theory to take this misalignment into account \cite{Zugarramurdi_2013_grating,Debiossac_PRB_2014,Debiossac_PRL_2014} whereas correcting data is not easy and restricted to simple systems~\cite{Debiossac_PRA_2014,Pollak_2015}.
The nature of the surface described by the diffracted intensities is unclear when these are recorded in inelastic conditions.
It seems difficult to avoid considering that the thermally averaged atomic position should be considered under deeply inelastic conditions, but the experiment suggests that this might not be the case in the quasi-elastic regime, where elastic diffraction is important.
The equivalence of the polar and azimuthal profiles, apparently valid in the quasi elastic regime, may not be valid when diffraction is completely inelastic with specific contributions from well localized trajectories inside the unit cell.
More experimental and theoretical work is needed to better model the inelastic diffraction in GIFAD.
There are obvious similarities with the multiphonon regime observed~\cite{Taleb_2017} and modelled~\cite{manson1991inelastic} in thermal energy atomic scattering, but these remain to be investigated in the GIFAD context.

\section{Acknowledgment} We are grateful to Hynd Remita for the irradiation of the LiF samples by $\gamma$ rays from the Cobalt source of the Institut de Chimie Physique at Orsay,  favoring further cleaving along (001) planes with large terraces and observation of elastic diffraction.
This work received support from LabEx PALM (ANR-10-LABX-0039-PALM) and Chinese Scholarship Council (CSC) Grant No. 201806180025.

\bibliography{bibliography}  
\end{document}